# Enhanced Slippery Behavior and Stability of Lubricating Fluid Infused Nanostructured Surfaces

*Reeta Pant,[a] Sanjeev Kumar Ujjain,[a] Arun Kumar Nagarajan[b] and Krishnacharya Khare[a]\**

[a]*Department of Physics, Indian Institute of Technology Kanpur, Kanpur - 208016, India*

[b]*Hindustan Unilever Research Centre, Bangalore - 560066, India*

E-mail: kcharya@iitk.ac.in

Keywords: Lubricants, Coating, Wetting, Adhesion, Liquid-liquid interface

ABSTRACT

Stability of lubricating fluid infused slippery surfaces is a concern for scientists and engineers and attempts are being made for its improvement. Lubricating oil coated slippery surface for aqueous drops is one of the important candidates in this class and their stability needs be improved to make them useful for practical applications. Cloaking of water drops with thin lubricant layer results in the loss of lubricant leading to deterioration of slippery behavior. Surface roughness or porosity provides larger surface area to the lubricating fluid and would to affect the stability of the lubricating film. Here we report the effect of surface roughness, from tens of nanometer to few microns, on the stability of slippery surface. Samples with small nanoscale roughness show improved performance in terms of contact angle hysteresis, critical tilt angle and slip velocity. Whereas large roughness samples show poorer performance compared to small nanoscale roughness and smooth samples. Small nanoscale roughness samples also show relatively slower deterioration against loss of lubricant during water flow. Once completely lost, the slippery behavior can be restored again simply by coating the sample again by the lubricating fluid.

**Introduction**

Superhydrophobic surfaces have been extensively used in last few years to achieve self-cleaning, anti-biofouling, anti-icing, liquid transportation etc. properties.[1-6] Fabrication of these surfaces is by mimicking the Lotus effect which requires creation of dual scale roughness along with coating of low surface energy material.[7-11] Superhydrophobic behavior of these surfaces is due to the composite surface made by the solid and the trapped air (Cassie-Baxter state).[12-14] Superhydrophobic behavior of these surfaces can be permanently destroyed if they experience vibration, pressure etc. due to metastable Cassie state.[15-17] Once destroyed, such surfaces show very poor water repellant and transportation behavior. This difficulty can be prevented if a water immiscible liquid (e.g. oil) is infused in the rough or porous structures where the infused liquid (oil) behaves as a lubricating film for water drops to slip. Nature presents such lubricant infused rough slippery surfaces in the form of Nepenthes pitcher plants where a thin lubricating water layer is present on the rough plant surfaces.[18] Such structures gained immediate attention of researchers to fabricate lubricating fluid infused slippery surfaces to slip test liquid drops. In lubricating fluid infused slippery surfaces, a low surface tension lubricating fluid is infused in textured or porous substrates and slippery behavior of various immiscible liquid drops is studied.[19-31] Slippery behavior on these lubricated surfaces is analyzed in terms of their slip velocity, contact angle hysteresis and critical tilt angle (minimum tilt angle at which test drop starts slipping). These slippery surfaces have found numerous applications in anti-icing, anti-corrosion, self-cleaning, self-healing, omniphobic behavior to name a few.[28, 32-38] Fabrication of these lubricating fluid infused slippery surfaces requires fulfillment of the conditions: (1) substrate should be completely wet with lubricating fluid, (2) immiscibility of lubricating fluid with test liquid and (3) substrate should be more wetting for the lubricating fluid as compared to the test liquid.[21, 28, 31, 39]  Porous, patterned or textured substrates have primarily been used to fabricate lubricant infused slippery surfaces. Such rough surfaces provide better adhesion to a lubricating fluid due to larger surface area compared to smooth surfaces. Aizenberg et al. used porous PTFE membranes and lithographically patterned post array with average roughness from 200 nm up to 2 μm to fabricate slippery surfaces.[28] Varanasi et al. also used lithographically patterned silicon micro post surfaces of 10 μm size to study drop mobility on slippery surfaces.[25] Huang et al. used porous polyelectrolyte multilayers to study omniphobicity of slippery surfaces.[21] But the systematic study of the effect of surface roughness on the slippery behavior of lubricating fluid infused surfaces in terms of stability and longevity is still missing. Such study will provide us the importance of surface roughness in fabricating slippery surfaces. Aizenberg et al. demonstrated effect of length scale of the surface texture on the slippery behavior in terms of contact angle hysteresis and critical tilt angle.[22] They studied the effect of substrate roughness on the slippery behavior against shear forces of different magnitudes. They concluded that amount of lubricant present on the substrates is not the only deciding factor for slippery performance but the surface roughness plays a dominating role in determining slip velocity as well as stability of the surfaces. Varanasi et al. found that drops of test liquid on lubricating fluid coated slippery surfaces are cloaked with a thin layer of lubricating fluid if the spreading coefficient of oil on water is positive.[25, 40] Cloaking of drops as well as deformation of three phase contact line (kink formation) in the lubricating film has recently been observed directly using laser scanning confocal microscope.[41, 42] Therefore once these lubricating oil cloaked test drops slip, they slowly remove the lubricating fluid from the substrate surface. After slipping sufficiently large amount of test liquid, deterioration in slippery behavior is expected. Therefore using porous or rough substrates provide better stability of lubricant which has been the case with all previous studies.

In this present study, we demonstrate the effect of substrate roughness on slippery behavior with silicone oil and water as lubricating fluid and test liquid respectively. Roughness on silicon substrates was created by using various etching techniques. Systematic analysis of critical tilt angle and corresponding water drop velocity and contact angle hysteresis were studied on smooth as well as rough hydrophobic substrates. Lubricated substrates with surface roughness in few tens of nanometer showed the better stability and longevity compared to smooth and large roughness samples.

**Experimental Section**

Single side polished silicon (Si) surfaces (UniversityWafers, USA), with thin native oxide, were used as solid substrates. These substrates were cleaned with ultrasonicating in ethanol and acetone followed by piranha cleaning (1:1, $H_2SO_4$ : 30% $H_2O_2$) at 80°C for 20 min. Roughness on cleaned Si substrates was generated using different wet chemical etching methods depending on the roughness amplitude. Small scale roughness (tens of nanometer) was obtained by immersing the cleaned Si substrates in aqueous solution of $NaBF_4$ and $AgNO_3$ at 80°C for 2 hrs.[43, 44] Subsequently, the resulting samples were immersed in aqueous solution of HCl and $HNO_3$ (1:1:1) overnight to remove the deposited silver dendrites. This process resulted in roughness in nanometer range depending on the concentration of $NaBF_4$ and $AgNO_3$ mixture. For large scale roughness (hundreds of nanometer), electrochemical etching of the Si substrates was done in aqueous solution of KOH and Isopropyl alcohol (IPA). Anisotropic etching of Si with KOH resulted in micron sized tetrahedrons. Micron size rough surfaces were obtained by immersing the cleaned Si substrates in aqueous solution of 4 M KOH/IPA for overnight. This process resulted in dual scale roughness with micron size tetrahedrons having nanoscale features on it. Roughness of smooth and etched samples was characterized using an Atomic Force Microscope (AFM) and 3-D Optical Profilometry. Large scale morphologies and etching uniformity were also analyzed using Optical and Scanning Electron Microscope (SEM).

Silicone oil (Sigma Aldrich, kinematic viscosity ~ 370 cSt) was used as lubricating fluid and was dip coated on the smooth and rough Si substrates. The oil coated substrates were then kept vertically for 30 mins to get rid of excess lubricant. Lubricating film thickness was estimated by weight difference method which came out as 3 μm and was kept constant for all the experiments. Subsequently, the lubricant coated samples were annealed at 150°C for 90 min to hydrophobize the Si surface which is a necessary condition for stable slippery surfaces.[31, 45] Critical tilt angle for slipping is defined as the minimum tilt angle at which a test drop (10 μl volume) starts slipping. Slip velocity of 10 μl aqueous drop at 12° tilt angle and contact angle hysteresis was used to quantify the slippery behavior the substrates. Stability test of the lubricant coated slippery surfaces was performed by step wise dispensing 50 ml volume of water (of 10 μl drops) followed by measuring the slip velocity.

**Results and Discussion**

Figure 1 summarises SEM images of rough silicon substrates obtained by different etching techniques. These etching techniques and respective parameters were chosen in a manner such that the surface roughness could be varied all the way from few nanometers up to few microns. $NaBF_4$ etched silicon substrates show uniform roughness over large area and the magnitude of roughness was controlled by controlling the solution concentration. KOH etching results into nano/micro sized tetrahedron structures due to anisotropic etching of silicon along <100> plane. To have stable slippery surfaces, as per the stability criteria, following conditions need to be satisfied: (i) complete spreading of lubricating fluid on the substrate, (ii) immiscibility of lubricating fluid with test liquid, and (iii) non-wetting of test liquid with solid substrate.[28, 31] These criteria can also be stated in terms of spreading parameters as given below:

$$S_{os} = \gamma_{sa} - \gamma_{os} - \gamma_{oa} \tag{1}$$

$$S_{ws} = \gamma_{sa} - \gamma_{ws} - \gamma_{wa} \tag{2}$$

where $S_{os}$ and $S_{ws}$ are spreading parameters of oil on substrate and water on substrate and $\gamma$ is interfacial tension and $o$, $s$, $w$ and $a$ present oil, substrate, water and air phases respectively.

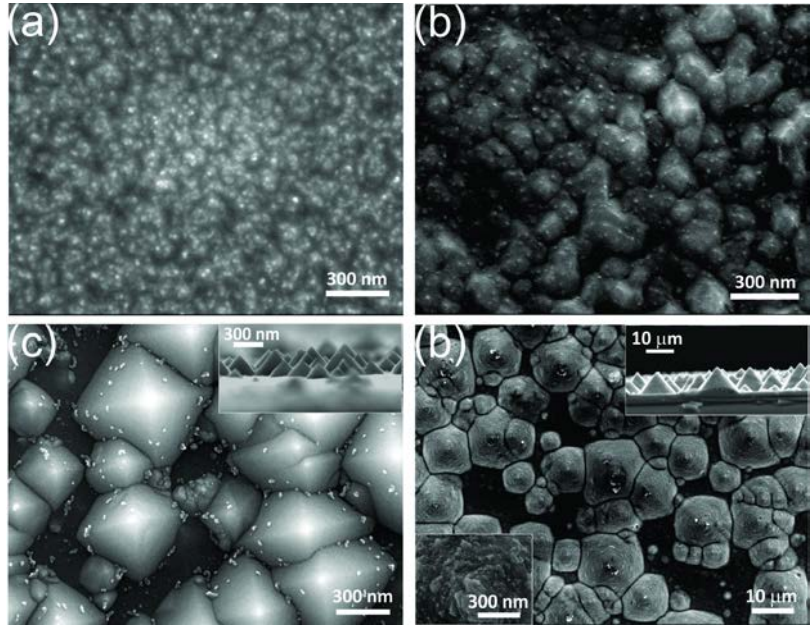

**Figure 1.** SEM images of etched Si substrates: a) 1 M NaBF$_4$ etched, b) 3 M NaBF$_4$ etched, c) electrochemically and d) overnight etched with KOH solution. Insets in images c) & d) show cross-sectional SEM showing tetrahedron structures after etching.

For hydrophilic (as-received) silicon substrates, values of $S_{os}$ and $S_{ws}$ results as zero and positive respectively which indicate complete spreading of both oil and water on the substrates. Whereas, as per the stability criteria, $S_{ws}$ should be negative indicating non-wetting of water on the substrates. This could be achieved by hydrophobiczing the surface by coating a self-assembled monolayer of silane/silicone molecules. Since we used silicone oil as lubricating fluid, annealing the oil coated substrates at 150°C for 90 mins made the silicon surface hydrophobic thus satisfied the stability criteria.[31, 45] To obtain thin lubricant film, dip coating was preferred over spin coating to have uniform lubricant thickness on smooth as well as rough surfaces. Subsequent to the lubricant coating, the samples were annealed at 150°C for 90 mins After coating these substrates with lubricating fluid, smooth as well as rough slippery surfaces were analyzed for their slippery characteristics and stability. Static contact angle of a sessile water drop was found to be 107° on all the samples (smooth and rough). This confirms that sessile water drop experiences a flat (smooth) lubricating fluid interface and depicts the equilibrium contact angle(Neumann's angle).[46] Table 1 summarizes the *rms* roughness of smooth and rough silicon substrates (measured by Atomic Force Microscope and 3D Optical Profilometer), contact angle hysteresis, critical tilt angle and the slip velocity on lubricating fluid coated substrates.

| Sample Name | Roughness method | *rms* roughness [nm] | CA hysteresis on slippery surface [°] | Critical tilt angle [°] | Slip velocity [mm sec$^{-1}$] |
|---|---|---|---|---|---|
| **S** | Smooth | 0.2 | 2.5 | 1 | 0.8 |
| **R1** | 1 M NaBF$_4$ etching | 24.5 | 1.5 | 1 | 1.0 |
| **R2** | 3 M NaBF$_4$ etching | 57.9 | 7 | 2 | 0.7 |
| **R3** | Electrochemical etching in KOH | 139.6 | 4 | 2 | 0.8 |
| **R4** | Overnight etching in KOH | 2900 | 15 | 10 | 0.3 |

**Table 1.** Comparison of smooth and rough silicon substrates for *rms* roughness, contact angle hysteresis, critical tilt angle and slip velocity.

Contact angle hysteresis was calculated by measuring water contact angles during advancing and receding drop volume cycle. Contact angle hysteresis is found to be the smallest in samples with small nano scale roughness (R1) whereas for smooth and other large roughness samples it was larger. This is because samples R1 have small nanoscale roughness providing better adhesion to the lubricating fluid thus enhancing the lubricant performance showing the lowest hysteresis. Whereas for large roughness samples, water drops feel underneath roughness which hinders the movement of the three phase contact point of water drops, thus increases the contact angle hysteresis. Similarly, critical tilt angle for slipping was also found to be the smallest on sample R1 among smooth and rough samples due to better lubricant performance on it. Samples with the largest roughness (R4) showed critical tilt angle of 10°, therefore all the slippery tests were performed at 12° tilt angle. Slip velocity of water drops is also found to be the largest on samples R1 compared to the other samples. So this study confirms that samples with small nanoscale roughness depict better slippery characteristics compared to smooth or large rough surfaces.

On smooth substrates it was observed that as more and more number of test drops is passed (slipped), slip velocity is found decreasing. This indicates that lubricant coated smooth slippery surfaces lose their slippery characteristics as a function of number of slipping drops (total aqueous volume). This deterioration in the slippery characteristics can be attributed to the removal of the lubricant. Varanasi et al. demonstrated that a thin layer of lubricating fluid cloaks a test aqueous drop if the spreading coefficient of lubricating fluid on water is positive. In our case, the value of this spreading coefficient $S_{ow} = \gamma_{wa} - \gamma_{wo} - \gamma_{oa}$ comes out as 8.5 mN/m which confirms the cloaking of water drops with lubricating fluid. Upon tilting, as these oil cloaked water drops slip, they also carry along that much volume of the oil with them. As more and more number of water drops are slipped through these surfaces, thickness of the lubricating fluid is decreased which results in deteriorated slippery performance. The slipped water drops were collected in a container where also we saw a thin oil layer floating which confirms the removal of oil with slipping water drops.

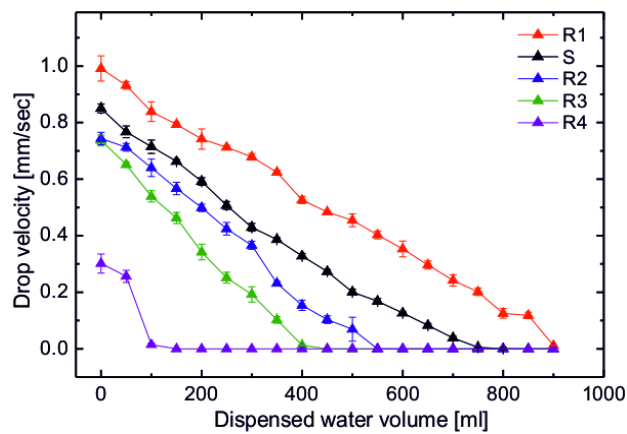

**Figure 2.** Stability test of lubricant on different rough surfaces as a function of dispensed water volume.

Figure 2 shows the plot of slip velocity of water drops on smooth as well as rough surfaces as a function of drop wise slipping water volume indicating deteriorated slippery performance. Water drops of 10 μl volume each were dispensed and slipped on smooth and rough slippery surfaces (with 12° tilt angle) for lubricant deterioration and slip velocity was measured after every 50 ml water volume is dispensed. The plot shows that approximately after 800 ml volume of water drops, the slip velocity becomes zero and the surface does not behave like slippery at all. Similar lubricant stability tests were performed on rough surfaces with increasing roughness. Samples R1, with 24.5 nm surface roughness, shows improved performance compared to smooth surfaces (S). Rate of deterioration was found to be smaller in samples R1. This superior performance is due to the improved adhesion of silicone oil with substrate because of the larger surface area of the rough samples. Upon further increasing the surface roughness, slippery performance starts deteriorating and the samples R2 behave slightly poorer than the smooth samples. Samples with higher roughness, R3 and R4, show even more poor performance and the total volume that can be slipped before complete deterioration decreases significantly. So overall, samples with small nanoscale roughness, show improved slippery behavior and better lubricant stability compared to smooth and large roughness samples. Smooth and large roughness samples show poor slippery characteristic with lower slip velocity and faster deterioration.

Stability of the slippery surfaces was also measured in terms of contact angle hysteresis for deteriorating samples. Figure 3 shows the contact angle hysteresis plot of smooth and rough samples as a function of dispensed water volume. For S and R1 samples, contact angle hysteresis always remained below 2° even after complete deterioration. This is because the surface roughness is extremely small and even very thin lubricating film can provide smooth movement of three phase contact point. For large roughness samples, hysteresis increases rapidly as more and more water volume is dispensed. This is because the lubricating film thickness decreases with dispensed water which exposes the roughness to slipping aqueous drops resulting in increased contact angle hysteresis and lower slip velocity. The last data point for different roughness samples in Figure 3 indicates the onset of complete deterioration where the slip velocity becomes zero. After this point, contact angle hysteresis cannot be defined as water drops completely get pinned. The accumulated data in the top of Figure 3 (above dashed line) indicates static contact angle on the deteriorating samples which corresponds to right Y-axis. This indicates that even though the samples have completely lost their slippery characteristic, they are still hydrophobic without any change in the static contact angle. This is because, after dip coating the lubricating fluid, samples were annealed at 150°C for 90 mins which made the sample surface hydrophobic. This hydrophobicity is not affected at all even if the samples lose their complete slippery characteristic.

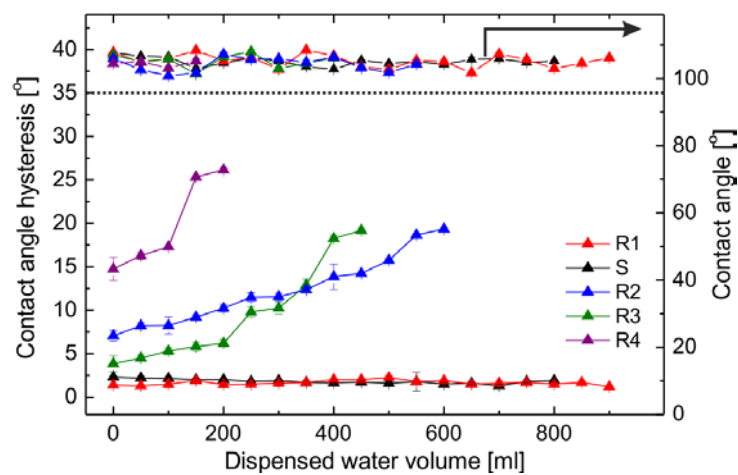

**Figure 3.** Stability test of lubricant on smooth and rough slippery surfaces in terms of contact angle hysteresis as a function of dispensed water volume (left Y-axis). The combined data above dotted horizontal line shows static contact angle (right Y-axis).

X-ray photoelectron spectroscopy (XPS) of bare, annealed and washed (after complete removal of the lubricant when the slip velocity becomes zero) samples is shown in Fig. 4. Initial survey scan confirmed the presence of Si(2p), C(1s) and O(1s) in all the three samples. Bare smooth and rough substrates show standard Si bonds (O-Si-O, Si-OH), C bonds (C-H, C-N, C-$O_x$) and O bonds (O-Si-O, C-$O_x$). Upon annealing, silicone molecules get covalently bonded with Si surface making it hydrophobic which is clear from the XPS spectrum with additional peaks of Si-O-Si, Si-C in Si(2p), C-Si in C(1s) and SI-O-Si in O(1s). These additional peaks confirm the covalent bonding nature of silicone molecule with Si surface upon annealing. After complete removal of the lubricant, when the slip velocity become zero and the contact angle hysteresis become maximum, XPS spectrum shows very similar characteristics as of annealed samples. This confirms that the even after complete removal of the lubricating fluid, at least a layer of covalently bonded silicone molecules is left at the surface. This was also confirmed by measuring the contact angle during lubricating fluid removal (Fig. 3 right Y axis, clubbed data points in above the dashed line).

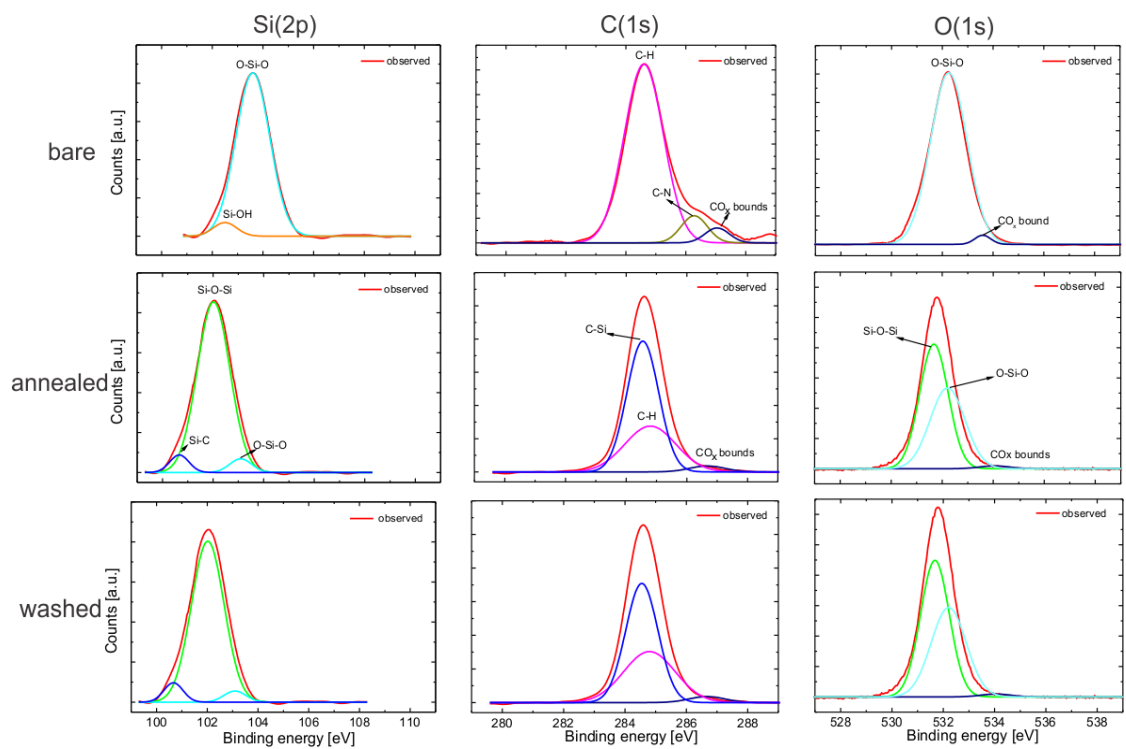

**Figure 4.** XPS spectrum of Si(2p), C(1s) and O(1s) of bare, annealed and completely washed samples.

As discussed earlier, due to positive spreading coefficient, water drops are cloaked with thin layer of lubricating oil. As these water drops slip, they remove that much amount of lubricating fluid thus thinning down the lubricating film thickness which is the main cause for the deterioration of the slippery surfaces. Water drop on lubricating oil film depict Neumann wetting regime where the water-oil-air three phase contact point gets deformed (kink formation) due to pulling of oil film by water drop.[46] Extent of this deformation (kink height) depends upon the lubricating film thickness.[47] For extremely thin oil films, the three phase deformation will be negligible and water drop will show Young's contact angle rather than Neumann's. We estimated the kink height of the deteriorating samples and found that as more and more water volume is dispensed through lubricant coated slippery surfaces, kink height is decreased. Figure 5 shows the snapshots of water drops on deteriorating R1 samples. It is clear from the snapshots that the kink height is decreased as the sample is deteriorated which confirms that the oil film thickness also decreases. The last snapshot corresponds to when the sample is completely deteriorated (slip velocity become zero and contact angle hysteresis become maximum) and the kink height is almost zero. This indicates that the lubricating oil thickness is almost zero and the drop contact angle is the Young's angle.

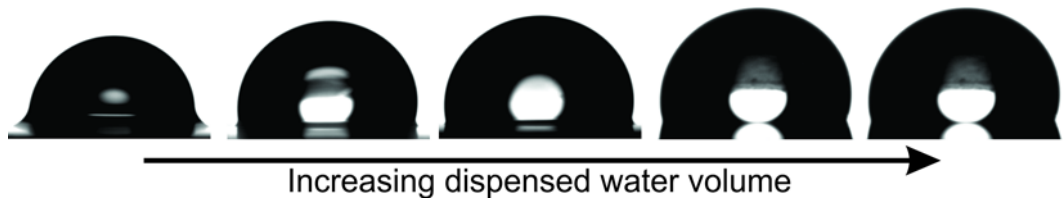

**Figure 5.** Snapshots of water drop on lubricant coated rough sample R1 with increasing dispensed water volume showing decrease in kink height.

We also studied the effect of slipping drop size (volume) on the deterioration of the slippery behavior. Since deterioration of slippery behavior is done by removal of lubricating oil by slipping drops, drop size is expected to play a major role in the longevity of slippery behavior. Figure 6 shows the slip velocity of water drops as a function of total dispensing volume for different size drops on smooth sample (S). It is clear from the experiment that less degradation in slippery behavior is observed with increasing drop volumes. This is because for given volume of water, larger size water drops will have smaller surface area compared to smaller size water drops. Therefore total volume of oil removed with larger drops of water will be smaller than compared to smaller drops of water. Therefore smaller water drops deteriorate the slippery behavior faster compared to bigger water drops. We also checked that if water is dispensed in form of very large drop or continuous film, no deterioration in slippery behavior is observed at all even after dispensing hundred litres of water. This is because no oil cloaking is possible for continuous dispensing water film resulting in no removal of oil thus showing no degradation in slippery behavior.

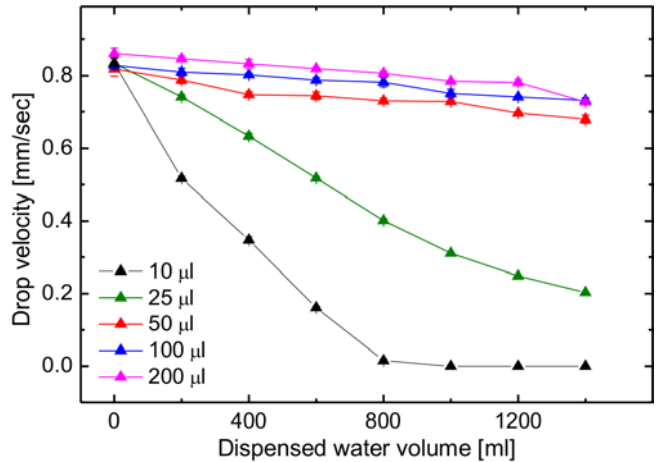

**Figure 6.** Stability of smooth slippery surface (S) as a function of total dispensed water volume for different individual volume of drops.

We also observed that coating these samples again with the lubricating fluid brings back their slippery characteristic. Complete deteriorated samples were recoated with silicone oil by dip coating and they again showed similar slippery characteristic with highest slip velocity and lowest contact angle hysteresis as if no deterioration happened to the sample. Therefore upon complete deterioration of slippery behavior due to removal or evaporation of the lubricant, they can be reused multiple times simply by recoating the lubricant again without any loss of the slippery characteristic.

**Conclusion**

We demonstrated that lubricating silicone oil coated silicon substrates, post annealing at 150°C for 90 mins, show good slippery characteristics for water drops. Due to positive spreading coefficient of the lubricating oil on water, the slipping water drops are cloaked with a thin layer of oil which is removed from the surface once these drops slip. This systematic removal of the lubricating oil from surface leads to deterioration of slippery behavior. To improve the performance (stability), rough substrates were used with surface roughness ranging from few nanometers up to few microns. It was found that surface with *rms* roughness around 25 nm significantly improves the slippery behavior in terms of contact angle hysteresis, critical tilt angle as slip velocity. This is because small nanoscale roughness samples provide better adhesion to the lubricating fluid thus showing improved slippery performance. Rough samples with larger roughness do not show any improvement in slippery behavior because large roughness starts affecting the drop motion increasing the contact angle hysteresis, and critical tilt angle and decreasing the slip velocity. We also observed that smaller sized water drops deteriorate slippery behavior faster compared to bigger sized drops due to large surface to volume ratio. If deteriorated, the slippery behavior can be recovered by coating the samples with the lubricating oil again.

**Acknowledgements**

This research work was supported by Hindustan Unilever Limited, India and DST, New Delhi through its Unit of Excellence on Soft Nanofabrication at IIT Kanpur.